\documentclass[%
reprint,
nofootinbib,
 amsmath,amssymb,
prl,
]{revtex4-2}
\usepackage{xcolor}
\usepackage{graphicx}
\usepackage{dcolumn}
\usepackage{bm}

\begin{document}

\preprint{APS/123-QED}

\title{Realization of the Pascal based on Argon using a Fabry-Pérot refractometer}


\author{Isak Silander$^1$}
\author{Johan Zakrisson$^{1,2}$}
\author{Ove Axner$^1$}
\author{Martin Zelan$^2$}
\email{martin.zelan@ri.se}
\affiliation{$^1$Department of Physics, Umeå University, 901 87 Umeå, Sweden \\
$^2$Measurement Science and Technology, RISE Research Institutes of Sweden, 501 15 Borås, Sweden
}



\date{\today}

\begin{abstract}
 
Based on a recent experimental determination of the static polarizability and a first-principles calculation of the frequency-dependent dipole polarizability of argon, this work presents, by use of a Fabry-Pérot refractometer operated at 1550 nm, a realization of the SI unit of pressure, the pascal, for pressures up to 100 kPa, with an uncertainty of [(0.98 mPa)$^2 + (5.8 \times 10^{-6} P)^2 + (26\times10^{-12}P^2)^2]^{1/2}$. The work also presents a value of the molar polarizability of N$_2$ at 1550 nm of 4.396572(26)$\times 10^{-6}$m$^{3}$/mol, which agrees well with previously determined ones in the literature. 

\end{abstract}

\maketitle

The implementation of the new SI in 2019 \cite{stock2019revision}, which fixed the value for the Boltzmann constant, has paved the way for innovative and alternative approaches to realize the pascal \cite{jousten2017perspectives}. Such methods, rooted in quantum properties, do not only hold the promise of surpassing the accuracy of traditional force-over-area techniques, (in particular in the lower pressure regions), they also offer the possibility for direct traceability to the SI, potentially eliminating the necessity for calibrations to other standards. 

While there exist various experimental techniques that utilize the link to the Boltzmann constant, the linchpin for all of them is accurate knowledge of the polarizability of the gas employed since its uncertainty directly constrains the uncertainty 
by which the pascal can be realized. In this context, the polarizability of He has been assessed by the use of \textit{ab initio} calculations with an uncertainty of 0.1 parts per million (ppm) \cite{puchalski2020qed}. Based on these first-principle calculations, a primary gas-pressure standard has been demonstrated by the use of dielectric-constant gas-thermometry (DCGT) with an expanded uncertainty (k=2) below 10 ppm in the 1 MPa to 7 MPa range, mainly limited by the knowledge of the viral coefficient and the compressibility of the capacitors \cite{gaiser2020primary}. 

An alternative approach is based on Fabry-Pérot (FP) refractometry. FP-refractometers are arguable the most sensitive of the quantum-based methods and can operate in lower pressure domains where conventional pressure standards have larger uncertainties. In short, the method relies on highly accurate measurements of the change in refractivity when gas is inserted in a high-finesse FP-cavity. This is done by measuring the change in frequency of laser light locked to a longitudinal mode of the cavity. From such a 
measurement, the molar density is calculated by the Lorentz-Lorenz equation, in which the knowledge of the molar polarizability is utilized. By using an equation of state and an accurate measurement of the gas temperature, the pressure can then be assessed.

In general, the main limiting factors when realizing pressure using FP-refractometers are the knowledge of the molar polarizability of the gas and the pressure induced cavity deformation of the cavity used. While using He as the measurement gas offers the lowest uncertainty of the molar polarizability (known from first-principle calculations) \cite{puchalski2020qed}, an accurate assessment of cavity deformation, which normally is performed by the two-gas method, requires a knowledge of the molar polarizability of  an additional gas \cite{stone2004using}.

To achieve the lowest uncertainty hitherto for realization of pressure using FP-refractometry, N$_2$ and He have been used by Egan et al.~to demonstrate a system operating at 633 nm with a claimed uncertainty of [(2~mPa)$^2 + (8.8 \times 10^{-6} P)^2]^{1/2}$ \cite{Egan2016} and by the authors at 1550 nm with an uncertainty of [(10~mPa)$^2 + (10 \times 10^{-6} P)^2]^{1/2}$ 
\cite{silander2021optical}. The latter system was recently upgraded yielding a pressure independent uncertainty of 0.75~mPa \cite{silander2023aninvar}.

In this work, by using Ar instead of N$_2$, and thus exploiting a recent low-uncertainty experimental determination of the static polarizability and a first-principles calculation of the frequency-dependent dipole polarizability of Ar \cite{gaiser2018polarizability,Lesiuk2023}, we present the so far most accurate realization of the pascal by a
quantum-based method. Based on this, we are also able to present a value of the molar polarizability of N$_2$ at 1550 nm at the gallium melting point (302.9146 K).  

The refractivity of the gas in the FP-refractometer is assessed using the expression \cite{silander2022situ}
\begin{equation}
        n-1  =\frac{\overline{\Delta\nu}+\overline{\Delta m}}{1-\overline{\Delta\nu}+\frac{\Theta_{G}}{\pi m_0} +  n \varepsilon}, 
    \label{eq:refractivity}
\end{equation} 
where $\overline{\Delta\nu}$ represents the relative frequency difference between an empty and filled cavity corrected for the phase shift of the light at the mirror surfaces and the Gouy phase, $\Theta_{G}$, $\overline{\Delta m}$ denotes the number of mode jumps performed during the filling (or emptying) of the cavity, and $\varepsilon$ is the refractivity-normalized pressure-induced relative cavity deformation, given by $\varepsilon_0 [1+\xi (n-1)]$, where, in turn, $\varepsilon_0$ is given by $\kappa \frac{2 R T}{3 A_R}$, where $\kappa$ is the deformation coefficient, $R$ is the molar gas constant, and $A_R$ is the molar polarizability, and $\xi$ is a coefficient that accounts for the non-linear behavior of pressure with respect to refractivity, given by a combination of the second refractivity and density virial coefficients, $B_R$ and $B_\rho(T)$, respectively, and $A_R$ \cite{Zakrisson2020}. For sub-atmospheric pressure, the molar density, $\rho$, can be calculated from the refractivity by use of the extended Lorentz-Lorenz equation,
    \begin{equation}
        \rho= \frac{2}{3A_R}(n-1)[1+ b_{n-1}(n-1)],
    \label{eq:rho}
    \end{equation}
where  $b_{n-1}$ is given by $-(1+4B_R/A_R^2)/6$, while the pressure can be assessed from the molar density and the temperature of the gas, $T$, as  
    \begin{equation}
        P = RT\rho [1+B_\rho(T) \rho].
    \label{eq:pressure}
    \end{equation}

\begin{figure}[htbp]
        \centering
        \includegraphics[width=\linewidth]{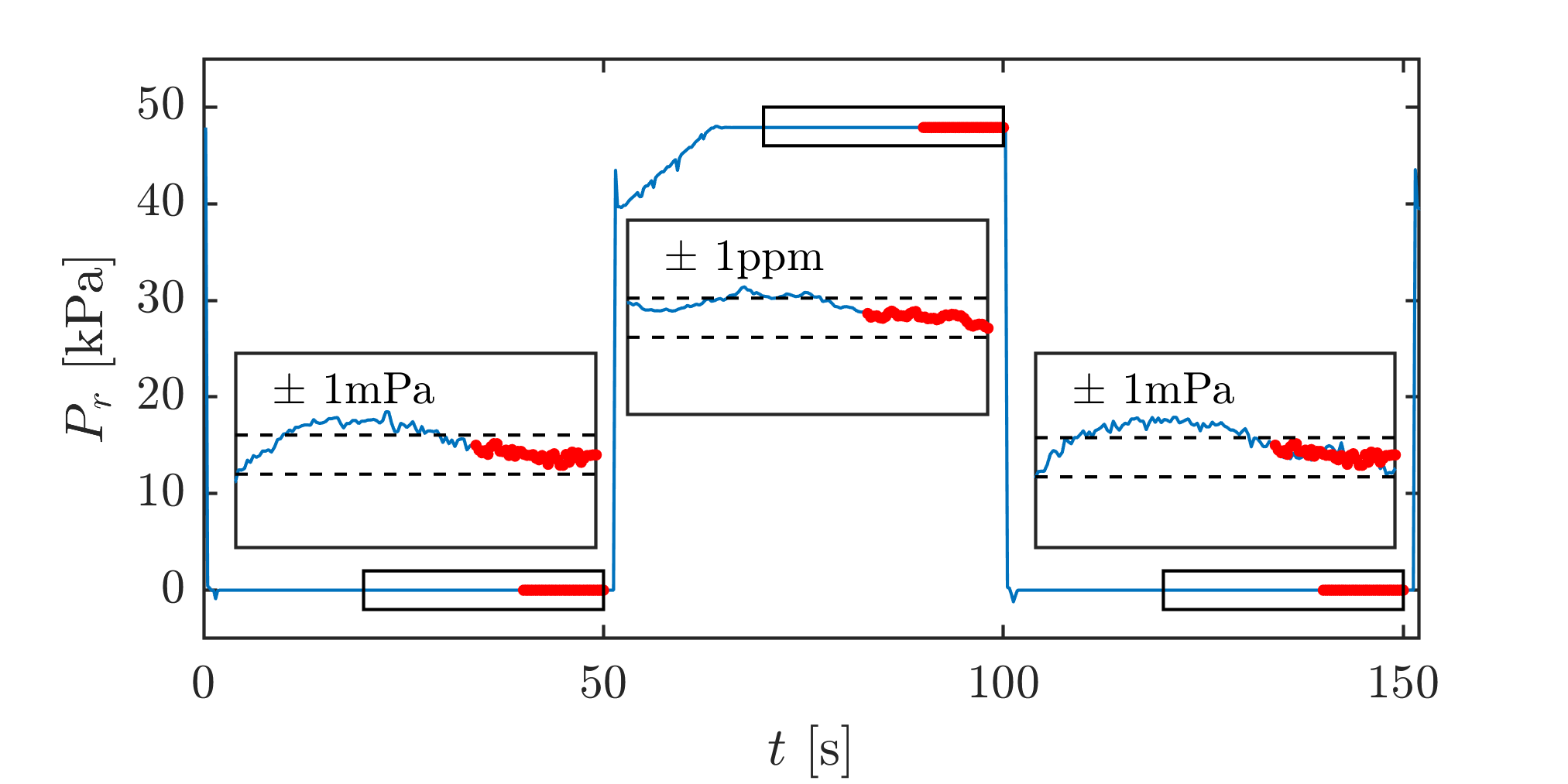}
        \caption{One measurement cycle from the refractometer connected to the DWPG with a set pressure to 47.89 kPa. The red markers indicate the samples that are averaged to evaluate the pressure. The middle insert represents a zoom in of data from a filled cavity measurement, while the first and the last inserts show data from two flanking empty cavity measurements. The dashed lines represent "typical" noise levels of 1 ppm and 1 mPa, respectively.
        }
        
       \label{fig:typical signal}
\end{figure}

The experimental setup has been described in detail in a multitude of publications, most recently in Silander et al.~\cite{silander2023aninvar}, but also in references therein. In short, the system consists of an Invar-based Dual FP cavity (DFPC) to which two narrow linewidth fiber-lasers are locked by Pound-Drever-Hall locking \cite{black2001introduction}. While one of the cavities acts as the measurement cavity in which gas is repeatedly filled and evacuated, the second one serves as the reference cavity to which the frequency is compared. The system is utilizing the Gas Modulation Refractometry (GAMOR) methodology \cite{silander2018gas}, which, as is shown in Fig.~\ref{fig:typical signal}, enables assessments of pressure with sub-ppm repeatability on the minute scale. 

The DFPC is connected to a DWPG that can set a pressure by applying a known force on a piston-cylinder ensemble with a known area \cite{sutton1994pressure}. In this work, the DWPG serves as a means to set a fixed pressure with high precision, so that the cavity deformation can be assessed with high accuracy, and as a conventional standard to which the results can be compared \cite{Zakrisson2020}. 

The DFPC and DWPG are connected by a gas supply system that can be swiftly switched between different gases. Before the measurements, the system underwent a leak/outgassing characterization to ensure a pure supply of gas.

The temperature of the cavity spacer was stabilized by a two-stage stabilization to within a few mK of the gallium-melting point (302.9146 K). To accurately assess the temperature of the cavity-spacer, and hence the gas \cite{rubin2022thermodynamic}, a Pt-100 sensor and a low-noise ohmmeter was used, which were compared to a gallium-melting point cell before and after each of the measurement series.

To properly evaluate the uncertainty of the DFPC-system, the uncertainty of each component that can have impact on the uncertainty of the instrumentation was assessed. While Ar is prophesied to yield the lowest uncertainty, and therefore is primarily targeted in this work, we have, for reasons of comparison, included also He and N$_2$. For the evaluations, the molar polarizability and the second refractivity- and density-virial coefficients with their corresponding uncertainties (\textit{k}=2) are needed for all gases addressed. The values of these entities were retrieved from the literature and were, in pertinent cases, recalculated to the wavelength and temperature used in this work. The values used are shown in Table \ref{tab:Gas constants}.
 
\begin{table}[htbp]
\centering
\caption{
Values for the gas coefficients used in this work for He, Ar, and N$_2$ at 302.9146 K and 1550.14 nm.}
\begin{tabular}{llr}
\hline
Coefficient & Value (uncertainty, \textit{k}=2) & Reference \\
\hline
Gas: He\\
\hline
$A_R$ & 0.51774512(10)$\times 10^{-6}$m$^{3}$/mol  & \cite{Puchalski2016} \\
$B_R$ & -0.05391(20) $\times 10^{-12}$m$^{6}$/mol$^2$ & \cite{Garberoglio2020}\\
$b_{n-1}$ & -0.03258(50) & \cite{Puchalski2016, Garberoglio2020} \\
$B_{\rho}$ & 11.90880(24)$\times 10^{-6}$m$^{3}$/mol & \cite{Czachorowski2020} \\
\hline
Gas: Ar\\
\hline
$A_R$ & 4.149661(22) $\times 10^{-6}$m$^{3}$/mol  & \cite{gaiser2018polarizability, Lesiuk2023} \\
$B_R$ & 1.71(11) $\times 10^{-12}$m$^{6}$/mol$^2$ & \cite{Garberoglio2020}\\
$b_{n-1}$ &-0.2330(43) & \cite{Garberoglio2020, Lesiuk2023} \\
$B_{\rho}$ &-14.565(54) $\times 10^{-6}$m$^{3}$/mol & \cite{egan2023optical} \\
\hline
Gas: N$_2$\\
\hline
$A_R$ & 4.396604(30) $\times 10^{-6}$m$^{3}$/mol  & \cite{egan2023optical, peck1966dispersion} \\
$B_R$ & 0.7411 $\times 10^{-12}$m$^{6}$/mol$^2$ & \cite{egan2023optical}\\
$b_{n-1}$ & -0.192                                  & \cite{egan2023optical} \\
$B_{\rho}$ & -3.920(92)$\times 10^{-6}$m$^{3}$/mol & \cite{egan2023optical} \\
\hline

\end{tabular}
  \label{tab:Gas constants}
\end{table}

The uncertainty 
from the entities that provide a pressure-independent contribution to the assessment of pressure, i.e.~the empty cavity repeatability, the residual pressure in the evacuated cavity, and gas contamination from leakages and outgassing, have recently been assessed for N$_2$ by Silander et al.~\cite{silander2023aninvar}. Following the  procedure in that work, these entities have here been assessed also for He and Ar. 

In addition to these components, a contribution from a previously omitted effect, which induces a change in length of the cavity, is included. It arises due to heat transfer from the mirrors to the gas, which effectively cools the mirrors that are heated due to absorbed laser light in the mirrors. As is mediated in a separate work \cite{Zakrisson2024b}, for pressures in the molecular region, for which the thermal conductivity of the gas is low, this effect is negligible. In the viscous region, on the other hand, it becomes non-negligible, in particular for He which has a relative large thermal conductivity and a low molar polarizability. This affects the length of the cavity, which, in turn, gives rise to an offset in pressure. By changing the laser power, it was found that for pressures in the viscous region, under the pertinent conditions (comprising a transmitted laser power of 1.3 mW and a cavity finesse of 10$^4$), this effect was found to be 354(23), 8.6(6), and 8.7(6) mPa for He, Ar, and N$_2$ respectively \cite{Zakrisson2024b}.

\begin{table}[htbp]
\centering
\caption{Uncertainty budget of the contributions to pressure assessments by use of the refractometer using He, Ar, and N$_2$ in terms of the expanded uncertainty ($\textit{k} = 2$) for individual modulation cycles of 100 s (50 s filling + 50 s evacuation).}
\begin{tabular}{p{14em}p{2.2em}p{2.2em}p{2.2em}p{2.2em}}

\hline
Components & He & Ar & N$_2$ & Type\\
\hline
\hline
\multicolumn{5}{l}{\textit{Constant terms} [$10^{-3}$/Pa]:}\\
Empty cavity rep., $\Delta f$           & 5.3       & 0.67      & 0.63      & A \\
Residual pressure, $P_0$                & 0.02      & 0.02      & 0.02      & A \\
Leaks and outgassing, $\delta(P_{leak})$     & 4.0       & 0.40      & 0.40      & A \\
Mirror cooling by gas
, $\Delta L$*       & 23        & 0.60         & 0.60         & A \\
\hline
\multicolumn{5}{l}{\textit{Linear terms} [$10^{-6}$]:} \\
Molar polarizability, $A_R$              & 0.2      & 5.4      & 6.8    & B \\
Temperature assessment, $T$              & 2.0      & 2.0      & 2.0    & A \\
Cavity deformation, $\varepsilon$        & 6.4      & 0.8      & 0.8    & A \\
Gas purity, $A_R$                        & 0.4      & 0.3      & 0.6    & B \\
Laser frequency, $\nu_0$                 & 0.2      & 0.2      & 0.2    & A \\
Gas heating, $T$                         & 0.3     & 0.3     & 0.3    & A \\
Penetration depth, $\gamma_s$            & 0.16     & 0.16     & 0.16   & A \\
Gouy phase, $\Theta_G$                  & 0.01     & 0.01     & 0.01   & B \\
\hline
\multicolumn{5}{l}{\textit{Quadratic terms}  [$10^{-12}$/Pa$^{-1}$]:} \\
$B_\rho(T)$                             & 0.15    & 21    & 37           &  B \\
$b_{n-1}$                               & 0.10    & 11    & 0**          & B \\
PV work, $T$                            & 10      & 10    & 10          & A \\
\hline
\textit{Total $U(P_{R})$}:*** &  &  \\
\multicolumn{5}{l}{He: $[(24$ mPa$)^{2}+(6.7\times10^{-6}P)^{2} + (10\times10^{-12}P^2)^2 ]^{1/2}$}\\
\multicolumn{5}{l}{Ar: $[(0.98$ mPa$)^{2}+(5.8\times10^{-6}P)^{2}+(26\times10^{-12}P^2)^2 ]^{1/2}$}\\
\multicolumn{5}{l}{ N$_2$: $[(0.96$ mPa$)^{2}+(7.2\times10^{-6}P)^{2}+(38\times10^{-12}P^2)^2 ]^{1/2}$}\\

\hline

\end{tabular}
\begin{tabular}{p{25em}}
\footnotesize

*~ It is important to note that this component, although being pressure-independent, exists only for pressures in the viscous region; it is negligible for pressures in the molecular region. In the intermediate regime its influence is non-trivial to estimate and its uncertainty can therefore differ slightly from what is given in this work, in particular for He \cite{Zakrisson2024b}.

**~ In reference \cite{egan2023optical}, measurement uncertainty was applied only to the second density virial coefficient while the second refractivity virial coefficient was given without any uncertainty.

***~ The expressions for the total uncertainty are given for gas in the viscous regime. In the molecular regime, the constant terms are 6.6, 0.78, and 0.75 mPa, respectively \cite{silander2023aninvar}.

\label{tab:Uncertety budget}

\end{tabular}
\end{table}

The uncertainty contributions from the pressure dependent part arises from a multitude of effects and is dominated by the molar polarizability, the cavity deformation, the gas temperature, and, for pressures approaching 100 kPa, the  density viral coefficients.


The expanded uncertainty of the temperature assessment of the cavity spacer
, assessed by the Pt-100, was estimated to 0.6 mK (2 ppm). To relate the temperature of the gas to the thermodynamic temperature two corrections were applied. Firstly, by comparing our gallium fixed point cell to that at RISE, the difference between the ITS90 scale and our fix point cell was assessed to 0.3 mK. Secondly, following Gaiser et al.~\cite{Gaiser2022}, the measured temperature was converted from the ITS90 scale to thermodynamic temperature by an amount of 3.86 mK.
The total uncertainty in the temperature assessment of 0.6 mK (k=2) was dominated by the uncertainty from the gallium fixed point cell at RISE.


To assess the pressure-induced cavity deformation, $\varepsilon$, an extended two-gas method based on Zakrisson et al.~\cite{Zakrisson2020} was utilized.  To account for the non-linear behavior of pressure with respect to refractivity, $\varepsilon_0$ was retrieved from the slope of second order fits to the difference in pressure, $\Delta P$, between the pressure set by the DWPG, $P_{set}$, and 
that assessed by the FP-refractometer.
From this, $\varepsilon_0$  was determined to 3076 ppm for He with an uncertainty of 6.4 ppm, which translates to an uncertainty of 0.80 and 0.75 ppm when Ar and N$_2$ are used as the measurement gas, respectively.

As the refractivity of a gas depends on its composition, it is necessary that the gases used have sufficient purity. Both the Ar and the N$_2$ were taken from the central gas supply at Umeå University, which guarantees at least 5N purity. The two dominating impurities are water and oxygen, both stated to be below 2 ppm. From this, we estimate the total contribution of these impurities to be at most 0.3 ppm for Ar and 0.6 ppm for N$_2$. 

For He, a 7N bottle was used in combination with a liquid N$_2$ trap, providing an estimated maximum uncertainty of 0.4 ppm.

The measurements of the laser frequency, $\nu_0$, were performed with an uncertainty of 0.2 ppm following the procedure described in \cite{Zakrisson2024a}.

In the previous characterisation of the system \cite{silander2021optical}, three processes that potentially can affect the system were omitted, viz.~the heating of the gas from the laser light  absorbed in the mirrors, the penetration  
of light into the coating of the mirrors (also referred to as phase-shift), and the Gouy phase. 

The heating of the gas due to losses in the cavity mirrors was assessed by changing the output power of the laser and measuring the change in the pressure assessed by the refractometer \cite{Zakrisson2024b}. It was found that it amounts to 1.2(1), 1.3(1), and 1.3(1) mK for He, Ar, and N$_2$, respectively.  

The uncertainties due to the penetration depth of the laser light into the mirror coating and the Gouy phase were estimated by the procedure given by Silander et al.~\cite{silander2022situ} to be 0.16 ppm and 0.01 ppm respectively.

In addition to the linear pressure dependent terms, the uncertainty is also influenced by the uncertainties in $B_\rho(T)$ and $b_{n-1}$. These will give rise to a quadratic pressure-dependence on the uncertainty. For this work, we used the uncertainty values that are given in Table \ref{tab:Gas constants}.

Finally, an additional quadratic term can potentially arise from PV-work when gas is let into and out from the cavity. No influence of this effect was detected in the characterization measurements. The uncertainty, ascribed to the PV-work, was taken as the noise of this measurement.

The uncertainty contributions from all the components described above are compiled in Table \ref{tab:Uncertety budget} where also the total uncertainty for He, Ar, and N$_2$ are presented.

Based on this results, it is possible to provide a value of $A_R$ for N$_2$ at 1550 nm. To do so, two measurements series of each of the gases were performed to which a second order function was fitted. The linear terms of these, which correspond to the proportional difference between the DWPG and the refractometer, were found to be -6.3(14), -6.4(18), and -13.8(12) ppm for He, Ar, and N$_2$ respectively. These differences can be used to update the value of the molar polarizability of N$_2$.

It was found that the combined uncertainty from the fits, the gas purities, the cavity deformation, and the molar polarizability became lower when the updating was made with respect to Ar than He, providing a value of $A_R$ for N$_2$ at 1550 nm of 4.396572(26)$\times 10^{-6}$m$^{3}$/mol. As is shown in Table \ref{tab:molarnitrogen}, this value agrees well with existing literature values and represents the hitherto most accurate value of the molar polarizability of N$_2$. 

\begin{table}[htbp]
\centering
\caption{The molar polarizability of N$_2$ at 1550.14 nm and 302.9146 K}
\begin{tabular}{llr}
\hline
Coefficients & Value (k=2) & Reference \\
\hline
$A_R$ & 4.396569(35) $\times 10^{-6}$m$^{3}$/mol &  \cite{Egan2016, peck1966dispersion} \\
 & 4.396604(30) $\times 10^{-6}$m$^{3}$/mol &  \cite{egan2023optical, peck1966dispersion} \\
& 4.396572(26) $\times 10^{-6}$m$^{3}$/mol &  this work \\
\hline

\end{tabular}
  \label{tab:molarnitrogen}
\end{table}

The quadratic terms retrieved from the same fits were found to be -23(17), -75(23), and -88(15) $\times$ $10^{-12}$ Pa$^{-1}$ for He, Ar, and N$_2$ respectively. This indicates a small, but statistically detectable non-linearity for He, and slightly larger non-linear discrepancies for Ar and N$_2$. While this might indicate slightly incorrect literature values of the refractive and density virials, it can alternatively be due to some yet unknown  effect in the DWPG used or the refractometer, suggesting that further investigation is needed. 
Irrespective of the origin of this discrepancy, it does though not significantly affect the claimed uncertainty of the pressure realization or the provided value of molar polarizability for N$_2$.



\begin{figure}[htbp]
        \centering
        \includegraphics[width=\linewidth]{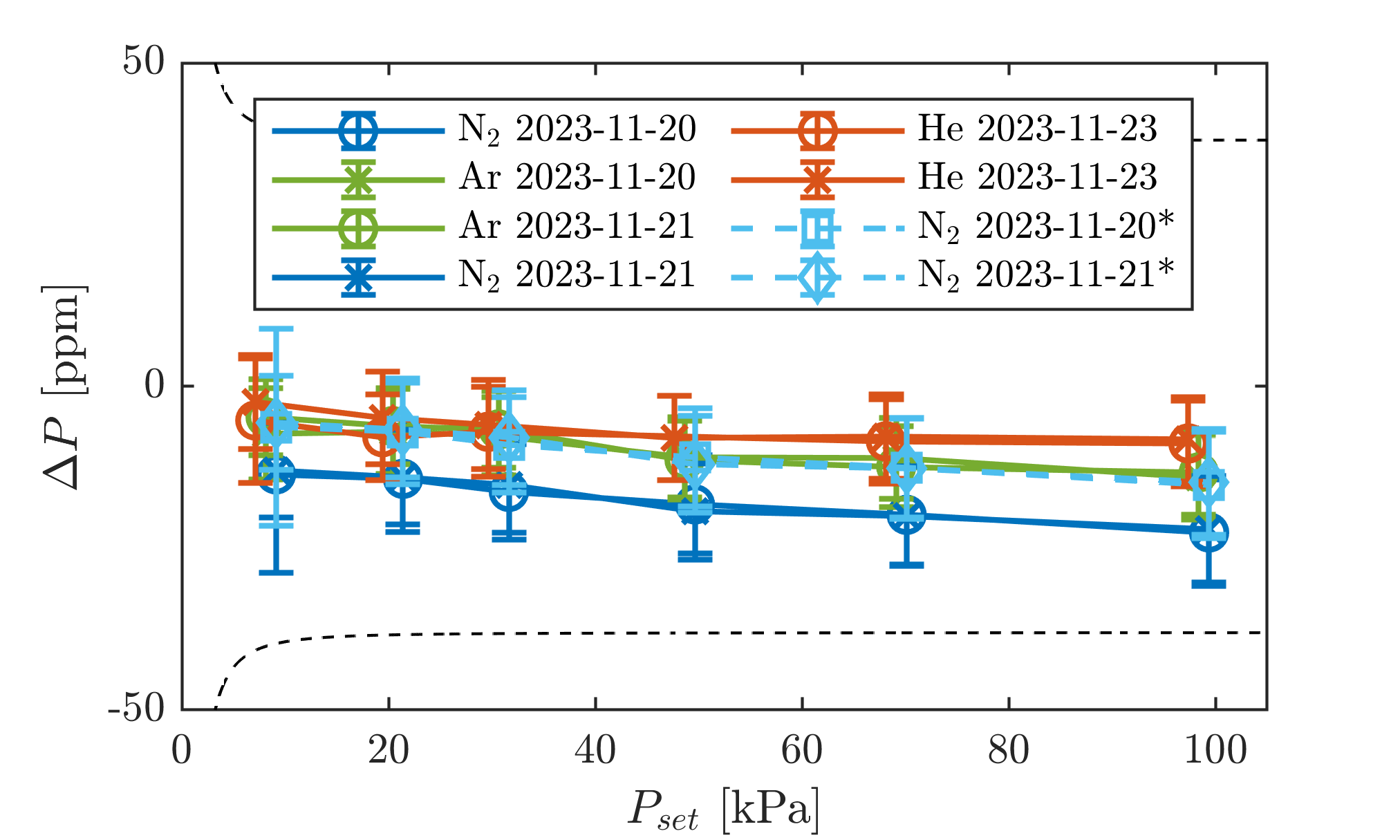}
        \caption{The difference between the assessed pressure by the DFPC and the set-pressure $P_{set}$ by the DWPG vs.~$P_{set}$ using Ar (green markers), He (orange markers), and N$_2$ (blue markers) as the measurement gas. The gas constants are taken from Table \ref{tab:Gas constants}, with the exception for the light blue data points marked with asterisks that use the updated value of $A_R$, 4.396572(26)$\times 10^{-6}$m$^{3}$/mol. The error bars represent the combined uncertainties (\textit{k} = 2) from the statistical spread and the uncertainty budget. The dashed black curves represent the uncertainty (\textit{k} = 2) of the DWPG.}
        \label{fig:vsdwpg}
\end{figure}

Finally, to verify the performance of the instrumentation, the data from the six measurement series are, in Fig.~\ref{fig:vsdwpg}, compared to the DWPG set-pressure. For N$_2$, two pairs of data sets are shown. The lower one, displayed in dark blue, shows the results when the $A_R$ value from the literature is used \cite{egan2023optical}, while the light blue data, with dashed lines, correspond to the updated value of $A_R$. All data points are well within the uncertainty of the DWPG. In addition, the refractivity measurements also agree with each other, demonstrating an excellent reproducible of the instrumentation. In the figure, the small non-linearities show up 
as small slopes in the response.



In conclusion, by addressing Ar, this work presents the so far most accurate realization of the pascal by a quantum-based method. Ar does not only have a lower influence of cavity deformation and is affected less by gas contamination than He, it has also a lower uncertainty of the molar polarizability than N$_2$. In addition, it has also been found that it does not suffer from such a large pressure offset due to heat transfer from the mirror to the gas as He does (for the pertinent instrumentation utilized in this work, 8.7 vs.~354 mPa \cite{Zakrisson2024b}).

Using Ar as the measuring gas, the system is capable of realizing the SI-unit for pressure in the viscous regime, for pressures up to 100 kPa, with an uncertainty of [(0.98 mPa)$^2 + (5.8 \times 10^{-6} P)^2 + (26\times10^{-12}P^2)^2]^{1/2}$. In the molecular regime, for which the mirror cooling due to transfer of heat to the gas is negligible, the uncertainty is even lower as the constant term drops to 0.78 mPa.

These results have also given a possibility to provide a value of the molar polarizability for N$_2$ at 1550 nm at the gallium melting point of 4.396572(26)$\times 10^{-6}$m$^{3}$/mol.

While the results presented in this work were made possible due to recently improved knowledge of the molar polarizability of Ar, it is important to emphasize that the limiting factor is still the molar polarizability. This is also the case for all other gas species, except for He. Consequently, improved values of the gas constants will therefore provide a direct improvement of the realization of the pascal. Of particular interest are first-principle calculations of the static polarizability of Ar with low uncertainties. This will not only improve on the accuracy, but it will also make the realization of the pascal fully independent of any conventional realization of pressure. 
 
\begin{acknowledgments}

This work was performed as a part of the "Metrology for quantum-based traceability of the pascal" project (MQB-Pascal 22IEM04), which has received funding from the European Partnership on Metrology, co-financed from the European Union’s Horizon Europe Research and Innovation Programme and by the Participating States. This research was additionally funded by Vetenskapsrådet (VR) grant number 621-2020-05105,  and the Vinnova Metrology Programme grant numbers 2022-02948. 

\end{acknowledgments}

\end{document}